# Magnetism and Electrical Conduction in Lightly-Doped Single-Layer High-$T_c$ Cuprate $La_2CuO_{4+\delta}$

Kota Miyakoshi,[1] Yuya Kinugawa,[1] Shusei Mizuta,[1] Yoshihiko Ihara,[1] Hiroyuki K. Yoshida,[1] Tohru Kurosawa,[2] Yasunori Toda,[2] Naoki Momono,[3] and Migaku Oda[1*]

*[1]Department of Physics, Hokkaido University, Sapporo 060-0810 Japan*

*[2]Department of Applied Physics, Hokkaido University, Sapporo 060-8628, Japan*

*[3]Department of Applied Sciences, Muroran Institute of Technology, Muroran 050-8585, Japan*

The temperature dependences of magnetization and electrical resistivity as well as their magnetic field dependences have been examined in lightly-doped single-layer cuprate $La_2CuO_{4+\delta}$ (LCO, hole-doping level $p(2\delta) \cong 0.03$) single crystals, in comparison with those in the extremely low doping region of $p \lesssim 0.015$ to uncover the intrinsic magnetism and electrical conduction of the Cu-O plane that exhibits both antiferromagnetic (AF) and superconducting (SC) orders simultaneously. In $p \cong 0.03$ SC LCO, the sub-lattice moments on Cu sites and their AF couplings are only $\sim 15\,\%$ smaller than those of the Mott-insulator parent material, suggesting that the localization of Cu$3d$ electrons remains very strong. Furthermore, we report that in the SC LCO, two-dimensional AF spin correlations develop rapidly from $T^* \cong 280$ K towards Néel temperature $T_N = 266$ K, where the out-of-plane resistivity starts to decrease largely. This might be responsible for the AF ordering at such a high temperature in the SC single-layer cuprate with $p \cong 0.03$.

## 1. Introduction

Nearly 40 years after the discovery of high critical-temperature ($T_c$) cuprate superconductors, the so-called "high-$T_c$ cuprates," the superconducting (SC) mechanism remains an intriguing problem studied extensively.[1)] Their parent materials are strongly correlated Mott-insulators, in which one electron of the Cu$3d_{x^2-y^2}$ orbital is localized on each Cu site due to strong on-site Coulomb interactions. The $S = 1/2$ spins of Cu$3d_{x^2-y^2}$ electrons are strongly coupled antiferromagnetically with superexchange interactions between nearest neighbor Cu sites. They begin to exhibit two-dimensional (2D) antiferromagnetic (AF) correlations within Cu-O planes at high temperatures over 1000 K corresponding to the in-plane AF coupling energy, develop with the lowering of temperature $T$, and become three-dimensionally ordered at around 300 K.[2,3)]

As holes are doped into Cu-O planes of the parent materials, the localization of Cu$3d$ electrons is weakened gradually. The AF order of Cu spins is suppressed rapidly with the increase of hole-doping level $p$ and, superconductivity, whose $T_c$ shows a dome-shaped $p$ dependence, appears around a certain value of $p$. Whether the AF and SC phases overlap in the $T$-$p$ phase diagram depends on the system.[4] Furthermore, around the optimal doping level $p_o$ ($0.16 - 0.18$ per $CuO_2$ square unit in most systems), where $T_c$ reaches a maximum, the magnitude of effective moments on Cu sites estimated in $La_{2-x}Sr_xCuO_4$ (LSCO) and $Bi_2Sr_2CaCu_2O_{8+\delta}$ (Bi2212) from the spin susceptibility, which is consistent with the $T$ dependence expected for $S = 1/2$ square-lattice Heisenberg antiferromagnets at high temperatures, is about one-fourth that of the parent material and, therefore, the localization of Cu$3d$ electrons is considered to be fairly weakened in the optimal electronic system for superconductivity.[5-7] On the other hand, the mechanism of high-$T_c$ superconductivity when the localization of Cu$3d$ electrons is very strong has also been proposed in some theoretical models such as the $t$-$J$ model.[8-10] It is therefore of great interest whether superconductivity occurs even in lightly-doped cuprates whose Cu$3d$ electrons are expected to be strongly localized. In such a hole-doping region, however, superconductivity was not found for a long time after the discovery of high-$T_c$ cuprates, giving rise to an interesting problem of whether superconductivity does not occur essentially in the electronic systems with strongly localized Cu$3d$ electrons or whether superconductivity is completely suppressed in the lightly-doped region from some reasons such as the susceptibility of lightly-doped electronic states to crystallographic disorders and the pronounced development of a gaplike structure around the Fermi level, the so-called "pseudogap" that competes with the superconductivity.[11-17]

In multi-layered high-$T_c$ cuprates $Ba_2Ca_{n-1}Cu_nO_{2n}(O,F)_2$ ($n = 3, 4$ and 5) which have $n$ Cu-O planes sandwiched by Ba-O(F) layers in the basic unit layer, it has been demonstrated by nuclear magnetic/quadrupole resonance (NMR/NQR) and angle-resolved photoemission spectroscopy (ARPES) measurements that superconductivity, coexisting with the AF order of Cu spins, occurs in the lightly-doped inner Cu-O planes where clean electronic systems can be realized with the screening effect of highly conductive outermost Cu-O planes on the disordered potentials from the charge-supplying Ba-O(F) layers.[4,18,19] This finding suggests that a small amount of holes can form SC pairs in the AF ordered state of localized Cu$3d$ electrons.

On the other hand, such microscopic coexistence of superconductivity and AF order has not been observed in most of single- and double-layered cuprates such as LSCO, $Bi_2Sr_2CuO_{6+\delta}$, $YBa_2Cu_3O_{6+\delta}$ and Bi2212. As is well-known, single-layer cuprate $La_2CuO_{4+\delta}$ (LCO) exhibits both AF and SC orders with the doping of holes at low concentrations due to introducing excess oxygen atoms.[20] However, this was considered to arise from the macroscopic phase separation into oxygen/hole poor and rich regions, whose $\delta$ values are located on the outer low and high sides of the first miscibility gap, respectively.[20-22] It was reported in previous studies that superconductivity occurs only in the oxygen/hole rich regions,[21,22] where excess oxygen atoms are intercalated into La-O layers with a certain period, forming a superlattice structure along the direction perpendicular to the layers or the so-called "stage structure".[23,24]

Very recently, it has been demonstrated in NQR measurements that superconductivity occurs in lightly-doped LCO with $\delta \cong 0.015$ or $p \cong 0.03$ (evaluating as $p = 2\delta$), which is of the so-called "infinite-stage structure" and does not undergo the phase separation, and coexists microscopically with the AF order.[25] The $^{139}$La-NQR spectrum for a $m = \pm\frac{5}{2} \leftrightarrow \pm\frac{7}{2}$ transition splits below $T_{\mathrm{N}} = 266$ K into very sharp two peaks, around which there are no spectral intensities arising from an oxygen/hole-rich phase as observed in NQR measurements on phase-separated samples with $\delta \cong 0.03$.[25,26] Furthermore, the $T$ dependence of nuclear spin-lattice relaxation rate $1/T_1$ measured with the NQR spectral peaks is consistent with the opening of a SC gap with lines of nodes below $T_{\mathrm{c}} = 32.5$ K.[25] These facts indicate that in lightly-doped LCO with $p \cong 0.03$, the superconductivity occurs in the AF ordered state that is formed homogeneously throughout the sample. Interestingly, in the coexistence state of superconductivity and AF order, the internal field on La sites due to the sub-lattice moments $M_{\mathrm{s}}$, originating from $3d_{x^2-y^2}$ electrons localized on Cu sites, is only $\sim 12\%$ smaller than that in the parent material, suggesting that Cu$3d$ electrons remain very strongly localized in AF-SC single-layer LCO with $p \cong 0.03$.[25]

In the parent material of LCO with $p = 0$, the Dzyaloshinskii-Moriya (DM) interaction is induced between nearest-neighbor Cu spins with the structural phase transition from tetragonal (I4/mmm) to orthorhombic (Cmca) at $T_{\mathrm{d}} \cong 530$ K and, in the AF ordered state, consequently, the sub-lattice moments on Cu sites are not completely parallel to the Cu-O basal plane or specifically the $c$-axis, one of the in-plane two axes ($a$- and $c$-axes) in the orthorhombic unit cell, but canted out of the basal plane (along the $b$-axis) by a tiny angle $\theta \cong 0.02°$.[27,28] The canted moments are ferromagnetically aligned in each Cu-O plane. Such weak ferromagnetic (WF) moments along the $b$-axis are inverted between adjacent Cu-O planes under zero magnetic field because of the very weak inter-plane AF couplings of Cu spins. The application of magnetic field $H$ higher than a critical value $H_{\mathrm{c}}$ along the $b$-axis induces a spin flop transition and aligns the WF moments throughout the crystal.[27-30] For $p \cong 0$, furthermore, the magnetic couplings between Cu spins have been evaluated with some experiments such as neutron inelastic scattering, two-magnon Raman scattering and magnetization measurements; the in-plane and inter-plane AF coupling constants, $J_{\parallel}$ and $J_{\perp}$, are $0.11-0.13$ eV and of the order of 1 μeV, respectively, and the DM coupling one $J_{\mathrm{DM}}$ is $\sim$1 meV.[5,23,27,29-34] According to the mean-field theory for strongly 2D AF spin systems, these magnetic couplings are crucial to determine the Néel temperature $T_{\mathrm{N}}$.[34] For the parent material of LCO, which is also that of the same type of single-layer cuprate LSCO, the value of $T_{\mathrm{N}}$ has been demonstrated to be $\sim$320 K with measurements on the $T$ dependence of magnetic susceptibility.[34] The AF order of Cu spins is affected by the doping of holes into Cu-O planes, as mentioned above. It is almost completely suppressed around $p \cong 0.015$ in the case of LSCO.[7] At doping levels lower than $p \cong 0.015$ in LCO, however, the value of $T_{\mathrm{N}}$ remains high, exceeding $\sim$300 K and, therefore, the magnetic properties seem to be little different from those of the parent material.[3,35,36] In contrast to the robust magnetism, the electrical conduction in LCO is strongly affected by holes doped into Cu-O planes even at extremely low concentrations, which are thermally transferred by 2D variable range hopping (VRH) among strongly Anderson-localized states.[30,37] Furthermore, it has been demonstrated that the conductivity of strongly Anderson-localized holes is significantly enhanced by the change in magnetic structure due to the field-induced WF transition.[27,30,38]

In the present study, we have examined the $T$ and $H$ dependences of magnetization, in-plane and out-of-plane resistivities in lightly-doped single-layer LCO single crystals with $p \cong 0.03$ where the superconductivity coexists microscopically with the AF order of Cu spins, in comparison with those in the extremely low doping region of $p \lesssim 0.015$. On the basis of obtained results, we discuss the magnetism and electrical conduction for the Cu-O plane that exhibits both AF and SC orders simultaneously. It has been demonstrated that in AF-SC LCO, the sub-lattice moments on Cu sites and their AF couplings ($M_{\mathrm{s}} \cong 0.71\mu_{\mathrm{B}}$ and $J_{\parallel} \cong 0.10$ eV) are only slightly smaller than those ($M_{\mathrm{s}} \cong 0.81\mu_{\mathrm{B}}$ and $J_{\parallel} \cong 0.12$ eV) of the Mott-insulator parent material, which arises from the still remaining strong localization of Cu3$d$ electrons. This finding confirms that the superconductivity whose $T_{\mathrm{c}}$ is as high as $\sim$30 K is caused by a small amount of conduction holes which are hosted in a robust AF ordered state of strongly localized Cu3$d$ electrons. Furthermore, it has been found that in AF-SC LCO, the development of 2D AF spin correlations accelerates from $T^{*} \cong 280$ K towards $T_{\mathrm{N}} = 266$ K, where the out-of-plane resistivity decreases largely while the in-plane one increases more rapidly. In light of this result, we will discuss the reason why the AF ordering occurs at such a high temperature exceeding 250 K in SC LCO with $p \cong 0.03$, whereas it is completely suppressed around $p \cong 0.02$ in the case of LSCO.

## 2. Experimental procedures

*2.1 Sample preparation and characterization*

Single crystals of LCO used in this study were grown in a 2.2 atm gas mixture of $20\ \%$ oxygen and $80\ \%$ nitrogen by the traveling solvent floating zone (TSFZ) method using a flux of $\mathrm{CuO}$ in the melting zone. Polycrystalline feed rods with a diameter of $\sim 4\ \mathrm{mm}$ were prepared from mixed powders of $\mathrm{La_2O_3}$ and $\mathrm{CuO}$ with a purity of 99.99%, whose composition satisfies the stoichiometric ratio for the constituent metal elements of LCO. The traveling rate of solvent was $\sim 0.6\ \mathrm{mm/hour}$. The present crystal growth method and conditions are similar to those conventionally employed for La-based high-$T_\mathrm{c}$ cuprates.[39,40]

Plate-shaped crystal pieces of several millimeters square cut out from the as-grown rod were annealed in a flowing Ar-gas at $\sim 900\ ^\circ\mathrm{C}$ for $\sim 100$ hours to remove a small amount of excess oxygen included in the growing process and obtain non-SC LCO samples with an extremely low hole-doping level. To prepare SC samples with a little bit larger amount of excess oxygen, the crystal pieces were further annealed in air at $\sim 600\ ^\circ\mathrm{C}$ for $5$ to $20$ hours after the Ar annealing. Furthermore, some as-grown crystal pieces were annealed in a vacuum of $\sim 10^{-3}$ torr at $\sim 900\ ^\circ\mathrm{C}$ for $1$ hour to obtain non-SC samples with $p \cong 0$, as well.

Figure 1 shows the $T$ dependences of magnetic susceptibility $\chi$ $(\equiv M/H)$ multiplied by $4\pi$ below $50$ K for two samples, referred to as LCO-B1 and LCO-C1, respectively. The data of magnetization $M$ were obtained in the process of raising temperature under a magnetic field $H$ of $2.0$ Oe parallel to Cu-O planes after cooling the sample at zero-field slowly from $300$ K to the lowest temperature examined and then confirming that $M$ exhibits an almost linear $H$ dependence at low fields around $2.0$ Oe. These samples are of exactly the same crystal piece but different in annealing process; the former was annealed only in Ar-gas and the latter was further annealed in air. LCO-B1 exhibits no sign of SC diamagnetism at least down to the lowest temperature (2 K) examined, whereas LCO-C1 exhibits a relatively sharp SC transition at $T_\mathrm{c} =$ $32.5$ K, as has been reported in our previous paper.[25] The magnitude of the low-$T$ diamagnetic susceptibility $\chi_\mathrm{cor}$ corrected for demagnetization field effects in LCO-C1 is very close to $1/4\pi$, indicating that the SC magnetic shielding of this sample is almost $100\%$. The sample LCO-C1 was also used in La-NQR measurements which have recently confirmed the microscopic coexistence of superconductivity and AF order in the single-layer cuprate.[25] The hole-doping level $p$ of LCO-C1 was estimated to be $\sim 0.03$ from the lattice constant $b$ at room temperature ($b = 13.145$ Å), revealed in X-ray diffraction measurements, by using its relation with the amount of excess oxygen $\delta$ or $p$ ($\equiv 2\delta$).[41] The $p$ value estimated in this way is consistent with the result obtained from the value of $T_N$ ($= 266$ K) by using its relation with $p$.[42] The $p$ value of LCO-B1 was estimated to be $\sim 0.015$ from the value of $T_N$ ($= 297$ K) similarly. Furthermore, for vacuum-annealed samples, the value of $T_N$ was slightly sample dependent, $317\ \text{K} - 323\ \text{K}$, giving $p \cong 0$ for these samples.[34,42] In this paper, A, B and C in sample names mean that their doping levels are $\sim 0$, $\sim 0.015$ and $\sim 0.03$, respectively, and different samples with the same doping level are distinguished with numbering, as shown above and in Table I for example.

For the evaluation of sample quality, we have examined microscopic crystallinity and electronic homogeneity from the sharpness of $^{139}$La-NQR spectral peaks for a $m = \pm\frac{5}{2} \leftrightarrow \pm\frac{7}{2}$ transition in the AF ordered state mainly. However, we have not yet evaluated the details of macroscopic crystallinity such as misorientation and twinning structure. The $^{139}$La-NQR spectral peaks for $p \cong 0.03$ are as sharp as those for $p \cong 0$; the full width at half-maximum (FWHM) of the peak intensity is only $\sim 20$ kHz in a wide $T$ range below $\sim T_N$ for each hole-doping level.[25] These results indicate that in the lightly-doped region of LCO, the microscopic properties such as the electric field gradient around La sites and the internal magnetic field on La sites due to the AF ordered Cu moments are extremely uniform, and the effects of disorder due to introducing a small amount of excess oxygen into La-O layers on the crystal periodicity and electronic homogeneity are very weak, which is contrast to the case of the substitution of Sr for La.[43] The FWHM of the $^{139}$La-NQR spectral peak at $T = 1.4$ K for $p \cong 0.025$ LSCO is $\sim 10$ times larger than that for $p \cong 0$.[43] Therefore, relatively clean Cu-O planes at a microscopic level are considered to be achieved in lightly-doped LCO crystals.

*2.2 Measurement methods*

Measurements on magnetization $M$ or magnetic susceptibility $\chi$ were carried out for air-, Ar- and vacuum-annealed samples by using a Quantum Design SC quantum interference device magnetometer, Magnetic Property Measurement System. In this study, $\chi$ was evaluated by dividing the measured value of $M$ by the applied field $H$, as already mentioned in Sec. 2.1. Measurements on electrical resistivity were carried out for air- and Ar-annealed samples with the four-probe method by using a Quantum Design Physical Property Measurement System. The electrodes, designed suitably for the four-probe measurements, were made by evaporating gold on the sample and then annealing it in an appropriate gas at $\sim 400\ ^{\circ}\mathrm{C}$ for $\sim 30\ \mathrm{min}$ to reduce their contact resistance; furthermore, thin gold lead wires with a thickness of $30\ \mu\mathrm{m}$ were attached on the electrodes with conductive silver paste. The $\sim 400\ ^{\circ}\mathrm{C}$ annealing for good electrical contacts was done, so as not to affect the sample quality, in the same gas as used in the higher-temperature annealing for the control of excess oxygen amount. In the present study, we measured the resistivity with flowing current along the in-plane Cu-O bond direction or the out-of-plane direction perpendicular to Cu-O planes; the former and latter resistivities are referred to as $\rho_{\parallel}$ and $\rho_{\perp}$, respectively.

## 3. Results

### *3.1 Magnetism in LCO*

Shown in Figs. 2(a) and 2(b) are the $T$ dependences of $\chi$ and $d\chi/dT$ which were examined in $p \cong 0.015$ (LCO-B1) and $0.03$ (LCO-C1) samples under a magnetic field of $H = 10\ \mathrm{kOe}$ applied along the out-of-plane direction, respectively. For $p \cong 0.015$, $\chi$ increases gradually down to $T_{\mathrm{N}} = 297\ \mathrm{K}$ with the lowering of $T$ and then decreases somewhat steeply with the three-dimensional (3D) AF ordering of Cu spins. Thio and Aharony have developed the mean-field theory on strongly 2D antiferromagnets such as LCO whose Cu spins are coupled by strong AF and DM interactions within Cu-O planes and by very weak AF interactions between adjacent Cu-O planes, and demonstrated that the increase of $\chi$ towards $T_{\mathrm{N}}$ is due to the gradual development of in-plane WF correlations of Cu spins caused by their strong AF and DM interactions, reflecting the $T$ dependence of 2D AF spin correlations.[7,27,34,44]

For $p \cong 0.03$, $\chi$ exhibits a very weak $T$ dependence down to $T^{*} \cong 280$ K, slightly higher than $T_{\mathrm{N}} = 266$ K, and increases steeply below $\sim T^{*}$ before turning to decrease at $T_{\mathrm{N}}$. This can be seen clearly in the $T$ dependence of $d\chi/dT$ (Fig. 2(b)), as well. Such a steep increase of $\chi$ near $T_{\mathrm{N}}$, reported previously for LCO samples with small values of $\delta$,[40,45] was understood in terms of the phase separation into oxygen/hole-rich and poor domains which exhibit SC and AF orders at low temperatures, respectively.[45] It was considered that as oxygen/hole-poor domains are formed partially below $\sim T^{*}$, 2D AF spin correlations rapidly develop there towards the 3D AF ordering at $T_{\mathrm{N}}$. This, combined with the DM interactions of Cu spins, leads to the rapid development of their in-plane WF correlations and consequently the steep increase of $\chi$. However, recent La-NQR measurements on the same $p \cong 0.03$ crystal have revealed that the superconductivity coexists microscopically with the AF order that is homogeneous throughout the sample; namely, it occurs without the phase separation, as mentioned in Sec.

1.[25] Therefore, the $T$ dependence of $\chi$ for $p \cong 0.03$ LCO suggests that the development of 2D AF spin correlations is gradual at temperatures higher than $\sim T^*$, but it accelerates from $\sim T^*$ towards $T_N$ throughout the sample. This is also considered to be a factor for the 3D AF ordering at such a high temperature exceeding $250$ K even for $p \cong 0.03$ in LCO, in addition to the fact that LCO has relatively cleaner Cu-O planes than LSCO, as discussed in Sec. 4.

Figure 3 shows the $H$ dependences of magnetization $M$, which were measured around $T_N$ under magnetic fields perpendicular to Cu-O planes in $p \cong 0$ (LCO-A1) and 0.03 (LCO-C1) samples. In these $H$ dependences of $M$, no hysteresis is observed between the field-increasing and decreasing processes. As seen in Figs. 3(c) and 3(d), for $T \gtrsim T^*$, the $H$ dependences of $M$ for $p \cong 0.03$ are almost linear at least up to 70 kOe, the highest field examined, and nearly fall onto a single straight line when the data for $M$ are normalized by its value at any specified field. For $T_N < T < T^*$, the magnetization normalized at a low field of 10 kOe follows the same straight line as for $T \gtrsim T^*$ in a $H$ region lower than $\sim$30 kOe, but gradually deviates downwards from the linear behavior with the increase of $H$, which becomes increasingly pronounced at lower temperatures (Fig. 3(d)). Such nonlinear $M$-$H$ curves are observed above $T_N = 320$ K for $p \cong 0$ (Figs. 3(a) and 3(b)), as reported in previous studies.[34] Notably, the nonlinearity in the $M$-$H$ curve remains prominent even at the highest temperature (350 K) examined, 30 K higher than $T_N$, for $p \cong 0$, while it becomes very weak above $\sim T^*$, 15 K higher than $T_N$, for $p \cong 0.03$. Thio and Aharony have demonstrated using the mean-field theory that the nonlinear behavior of magnetization can be explained in terms of the effect of very weak inter-plane AF couplings on Cu spins which are dominated by strong AF-WF correlations within Cu-O planes.[34] The inter-plane AF couplings of Cu spins have the effect of canceling out the WF moments between upper and lower AF-WF correlated regions within adjacent Cu-O planes. According to the mean-field theory, this effect becomes stronger nonlinearly with the increase of $H$, and significant in such a case that the 2D AF-WF spin correlations develop strongly even if the inter-plane AF couplings are very weak.[34] Therefore, the present observation on the nonlinear behavior of magnetization at $T_N < T < T^*$ for $p \cong 0.03$ LCO strongly supports the suggestion from $\chi$-$T$ measurements that the development of 2D AF-WF spin correlations will accelerate from $\sim T^*$ towards $T_N$.

In the AF ordered state below $T_N$, $M$ jumps at the critical field $H_c$ of field-induced WF transition, as shown in Figs. 3(a) and 3(c). The field-induced WF transition is very sharp except for a very narrow $T$ region near $T_N$ even in SC samples with $p \cong 0.03$. This fact also indicates that the AF ordered state is very homogeneous throughout the samples, as demonstrated in NQR measurements.[25] In Fig. 4(a), the magnitude of canted WF moment per Cu site, $M_{\mathrm{WF}}$, is plotted as a function of $T/T_{\mathrm{N}}$ for $p \cong 0.015$ ((LCO-B1) and 0.03 (LCO-C1) samples. This is estimated from the magnetization jump at the field-induced WF transition (see the inset of Fig. 4(a)) and represented in a unit of Bohr magneton $\mu_{\mathrm{B}}$. The data of $M_{\mathrm{WF}}(T)$ normalized with the value at $T \ll T_{\mathrm{N}}$, defined as $M_{\mathrm{WF}}(0)$, $M_{\mathrm{WF}}(T)/M_{\mathrm{WF}}(0)$, is also plotted as a function of $T/T_{\mathrm{N}}$ in Fig. 4(b), together with those of $H_{\mathrm{c}}(T)$. The value of $M_{\mathrm{WF}}(0)$, evaluated as an average of the data in a low-$T$ region ($T/T_{\mathrm{N}} < 0.35$) where they tend to be saturated, is $\sim 3.1 \times 10^{-3} \mu_{\mathrm{B}}$ for $p \cong 0.015$ and $\sim 2.9 \times 10^{-3} \mu_{\mathrm{B}}$ for $p \cong 0.03$. The $T$ dependence of $H_{\mathrm{c}}$ for $p \cong 0.03$ includes data determined from the $\rho$-$H$ curves, which exhibit comparatively sharp reductions with the field-induced WF transition, as well. When the data over the entire $T/T_{\mathrm{N}}$ range are normalized with $M_{\mathrm{WF}}(0)$ for each sample, they nearly fall on a single curve irrespective of hole-doping level, which resembles the result due to the mean-field approximation for $S = 1/2$ Heisenberg antiferromagnets, given by the Brillouin function for $S = 1/2$. Furthermore, the $T$ dependence of $H_{\mathrm{c}}$ is almost the same with that of $M_{\mathrm{WF}}$. For the low-$T$ ($T \ll T_{\mathrm{N}}$) value of $H_{\mathrm{c}}$, $H_{\mathrm{c}}(0)$, is $\sim 6.1 \times 10$ kOe for $p \cong 0.015$ and $\sim 5.9 \times 10$ kOe for $p \cong 0.03$. In the present study, it has also been revealed from the $M$-$H$ curve measured at $T = 50$ K ($\ll T_{\mathrm{N}}$) in a sample (LCO-A2) with $p \cong 0$ and $T_{\mathrm{N}} = 323$ K that $M_{\mathrm{WF}}(0) \cong 3.4 \times 10^{-3} \mu_B$ and $H_{\mathrm{c}}(0) \cong 6.2 \times 10$ kOe.

Using the magnitude of sub-lattice moments $M_{\mathrm{s}}$, their canting angle $\theta$, and the inter-plane AF coupling constant $J_{\perp}$ defined as per two Cu-Cu bonds,[34] the magnitude of canted WF moments $M_{\mathrm{WF}}$ and the critical field $H_{\mathrm{c}}$ of field-induced WF transition are given by the following equations:[27,30,34]

$$M_{\mathrm{WF}} = M_{\mathrm{s}}\theta, \tag{1}$$

$$H_{\mathrm{c}} = \frac{2J_{\perp}}{(g\mu_{\mathrm{B}})^2}\frac{M_{\mathrm{S}}}{\theta}. \tag{2}$$

Namely, the former and the latter are proportional to $M_{\mathrm{s}}\theta$ and $M_{\mathrm{s}}/\theta$, respectively. The canting angle $\theta$ is expressed as $\theta = J_{\mathrm{DM}}/2J_{\parallel}$,[27,34] and the DM coupling constant $J_{\mathrm{DM}}$ is given as follows:[27,34]

$$J_{\mathrm{DM}} \cong \phi J_{\parallel}\frac{\Delta g}{g}, \tag{3}$$

where $\phi$ is the rotation angle of $\mathrm{CuO_6}$ octahedra around the *a*-axis associated with the structural phase transition at $T_{\mathrm{d}}$ or the octahedral rotation order parameter, $g$ the $g$-value of $\mathrm{Cu}3d_{x^2-y^2}$ electrons, and $\Delta g \equiv g - 2$.

Therefore,

$$\theta \cong \phi \frac{\Delta g}{2g}, \quad (4)$$

and is nearly proportional to $\phi$. It has been demonstrated in neutron diffraction experiments on LCO that $\phi \propto (1 - T/T_\mathrm{d})^{0.238}$ and tends to be saturated at low temperatures below $T_\mathrm{N} \cong T_\mathrm{d}/2$.[46] This, together with the relation $\theta \propto \phi$, indicates that $\theta$ is only very weakly dependent on $T$. Consequently, the $T$ dependences of $M_\mathrm{WF}$ and $H_\mathrm{c}$ arise from that of $M_\mathrm{s}$, and are almost the same with each other, as observed in LCO-B1 and LCO-C1 (Fig. 4(b)). In fact, they are in good agreement with the experimental result on $M_\mathrm{s}$ obtained by the neutron scattering technique.[3,46]

It has recently been demonstrated in La-NQR measurements that the internal field on La sites at $T \ll T_\mathrm{N}$ in LCO-C1 with $p \cong 0.03$, reflecting the magnitude of sub-lattice moments $M_\mathrm{s}(0)$, is $\sim 12\%$ smaller than that in LCO with $p \cong 0$, which corresponds to $\sim 4p$.[25] Here, we assume that the $p$ dependence of $M_\mathrm{s}$ at $T \ll T_\mathrm{N}$, $M_\mathrm{s}(0,p)$, can be written as $M_\mathrm{s}(0,p) \cong M_\mathrm{s}(0,0)(1-4p)$ at low hole-doping levels. Interestingly, the data of $M_\mathrm{WF}$ at $T \ll T_\mathrm{N}$, $M_\mathrm{WF}(0,p)$, for $p \cong 0$, $0.015$ and $0.03$ can be fitted to the same function of $p$ as for $M_\mathrm{s}$ with a scale factor. Namely, $M_\mathrm{WF}(0,p)$ is expressed as $M_\mathrm{WF}(0,p) \cong (3.4 \times 10^{-3}\mu_\mathrm{B})(1-4p) \cong (3.4 \times 10^{-3}\mu_\mathrm{B})\, M_\mathrm{s}(0,p)/M_\mathrm{s}(0,0)$. Therefore, one can see from eq. (1) that $\theta \cong (3.4 \times 10^{-3}\mu_\mathrm{B})/M_\mathrm{s}(0,0)$, being little dependent on $p$. Furthermore, this result for $\theta$ is obtained directly from the $p$ dependence of $\phi$ by using the relation between $\theta$ and $\phi$ (eq. (4)), as shown in the next paragraph. This finding indicates that the present assumption $M_\mathrm{s}(0,p) \cong M_\mathrm{s}(0,0)(1-4p)$ is an appropriate approximation for the $p$ dependence of $M_\mathrm{s}$ at $T \ll T_\mathrm{N}$ in lightly-doped LCO samples.

Neutron diffraction experiments on LCO have reported that the octahedral rotation order parameter $\phi$ has the following relation with the orthorhombic distortion parameter $r$, defined as $r \equiv 200\,(c-a)/(a+c)$: $\phi \propto r^{1/3}$,[46] and the $\delta$ or $p$ dependence of $r$ at $T = 10$ K is expressed as $r \cong 1.7 - 10\delta$ or $1.7 - 5p$.[47] Hence, $\phi$ at low temperatures is proportional to $(1.7-5p)^{1/3} \propto 1 - 0.98p$ for small values of $p$. Using the expression $\phi(p) \cong \phi(0)(1-0.98p)$, where $\phi(0) \cong 0.092$ ($\sim 5°$) as reported from neutron diffraction structural analyses at $T = 10$ K in LCO samples with $p \cong 0$,[21,48] one can obtain $\phi \cong 0.091$ for $p \cong 0.015$ and $\phi \cong 0.090$ for $p \cong 0.03$ in addition to the value of $\phi$ for $p \cong 0$; namely, $\phi$ is almost the same in the lightly-doped region. Furthermore, the $g$ or $\Delta g$ value, related to the transition of a hole in the crystal-field-split $3d_{x^2-y^2}$ orbital to the other $3d$ orbitals due to spin-orbit coupling for $Cu^{2+}$, is considered to be little different among these samples. Some values of $g$ around 2.2 have been reported so far for LCO.[27,34,49,50] Here, we take 2.2 as the $g$-value, which has been used to evaluate some quantities for the magnetic properties of LCO and provided reasonable results for them.[27,34,51] Thus, one can obtain the values of $\theta$ for the three doping levels from the relation between $\theta$ and $\phi$ (eq. (4)), and demonstrate that $\theta$ is little dependent on $p$, which is $\sim 0.0042$ ($\sim 0.24°$) for $p \cong 0$ and $\sim 0.0041$ ($\sim 0.23°$) for $p \cong 0.015$ and $0.03$.

By using the relation $M_s(0) = M_{WF}(0)/\theta$ and the values of $M_{WF}(0)$ and $\theta$ obtained in this study, $M_s(0)$ can be evaluated to be $\sim 0.81\,\mu_B$, $\sim 0.75\,\mu_B$ and $\sim 0.71\,\mu_B$ for $p \cong 0$, $0.015$ and $0.03$, respectively. These values for $M_s(0)$ follow its $p$ dependence assumed in this study, $M_s(0,p) \cong 0.81\mu_B(1-4p)$, consistently. Furthermore, it is noteworthy that the low-$T$ sub-lattice moments for the parent material of LCO is as large as $\sim 70\%$ of the full magnetic moments expected for $S = 1/2$ and $g = 2.2$, which is in good agreement with the renormalization factor due to strong quantum fluctuations in $S = 1/2$ square-lattice antiferromagnets.[27,52-54]

The inter-plane AF coupling constant $J_\perp$ is expressed as $J_\perp = H_c M_{WF} / \left\{2\left(\frac{M_s}{g\mu_B}\right)^2\right\}$ from eqs. (1) and (2). Employing this expression and the values of $H_c(0)$, $M_{WF}(0)$ and $M_s(0)$ for the three hole-doping levels, one can evaluate $J_\perp$ to be ~4.5 $\mu$eV for $p \cong 0$ and $\sim 4.7\,\mu$eV for $p \cong 0.015$ and $0.03$; it is little dependent on $p$ in the lightly-doped region.

In the mean-field theory by Thio and Aharony on the strongly 2D AF spin system of lightly-doped LCO in which Cu spins are coupled by in-plane DM interaction $J_{DM}$ and very weak inter-plane AF interaction $J_\perp$ in addition to strong in-plane AF interaction $J_\parallel$, the Néel temperature $T_N$ satisfies the following equation:[34]

$$\frac{1}{\chi_{2D}^{\dagger}(T_N)} = 4\theta J_{DM} + J_\perp. \tag{5}$$

In eq. (5), $\chi_{2D}^{\dagger}(T_N)$ is the staggered susceptibility at $T_N$ and given for $S = 1/2$ 2D Heisenberg antiferromagnets as follows:[27,28,55]

$$\chi_{2D}^{\dagger}(T_N) = \frac{\left(\frac{\xi_{2D}(T_N)}{a_0}\right)^2}{k_B T_N}, \tag{6}$$

where $\xi_{2D}(T_N)$ and $a_0$ are the 2D AF spin correlation length at $T_N$ and the lattice constant (3.8 Å for the square-lattice Cu-spin system of LCO). Thus, eq. (5) is rewritten as eq. (7).

$$k_B T_N = (4\theta J_{DM} + J_\perp)\left(\frac{\xi_{2D}(T_N)}{a_0}\right)^2. \tag{7}$$

Furthermore, $\xi_{2D}(T)/a_0$ is given as $\xi_{2D}(T)/a_0 = C_\xi \exp(2\pi\rho_S/k_B T)$, where $C_\xi$ is a constant value and $2\pi\rho_S$ is a spin stiffness constant given by $\kappa J_\parallel$ with a factor $\kappa$ of order unity.[55] The values of $C_\xi$ and $\kappa$ have been calculated to be $\sim 0.27$ and $\sim 1.15$ for $S = 1/2$ square-lattice Heisenberg antiferromagnets, respectively,[53-55] which are considered to be reasonable for the Cu-spin system of lightly-doped LCO from the following reason: the above expression of $\xi_{2\mathrm{D}}(T)$ for $C_\xi \cong 0.27$ and $\kappa \cong 1.15$ can almost reproduce the results obtained in neutron inelastic scattering experiments at high temperatures in lightly-doped LCO crystals with $T_\mathrm{N} \gtrsim 200$ K.[31,55] In this study, also adopting the above values for $C_\xi$ and $\kappa$, the term relating to the correlation length in eq. (7), $\xi_{2\mathrm{D}}(T_\mathrm{N})/a_0$, is expressed as follows:

$$\frac{\xi_{2\mathrm{D}}(T_\mathrm{N})}{a_0} \cong 0.27\exp(1.15 J_\parallel / k_\mathrm{B} T_\mathrm{N}). \quad (8)$$

Employing eqs. (3), (4) and (8), eq. (7) can be approximately rewritten as

$$\frac{k_\mathrm{B} T_\mathrm{N}}{(0.27)^2} \cong \left\{2\phi^2 \left(\frac{\Delta g}{g}\right)^2 J_\parallel + J_\perp\right\} \exp\left(\frac{2.3 J_\parallel}{k_\mathrm{B} T_\mathrm{N}}\right). \quad (9)$$

Here, defining $\alpha$, $\beta$, $\gamma$ and $\zeta$ as $\alpha \equiv 2\phi^2(\Delta g/g)^2$, $\beta \equiv J_\perp$, $\gamma \equiv 2.3/k_\mathrm{B} T_\mathrm{N}$ and $\zeta \equiv k_\mathrm{B} T_\mathrm{N}/(0.27)^2$, eq. (9) is expressed as $\zeta \cong (\alpha J_\parallel + \beta)\exp(\gamma J_\parallel)$. This is an equation with $J_\parallel$ as the unknown, because the values of physical quantities and constants included in $\alpha$, $\beta$, $\gamma$ and $\zeta$ have already been clarified for the LCO samples studied here. The solution for $J_\parallel$ is given as eq. (10) by using the Lambert's W-function:

$$J_\parallel \cong \frac{1}{\gamma} W\left(\frac{\gamma\zeta}{\alpha} \exp\left(\frac{\beta\gamma}{\alpha}\right)\right) - \frac{\beta}{\alpha}. \quad (10)$$

Thus, the in-plane AF coupling constant $J_\parallel$ can be evaluated from eq. (10) for each sample; it is $\sim 0.11$ eV for $p \cong 0.015$ and $\sim 0.10$ eV for $p \cong 0.03$. Regarding $J_\parallel$ for $p \cong 0$, one can also obtain its value to be $\sim 0.12$ eV, which is consistent with previous reports such as neutron scattering, Raman scattering and magnetic susceptibility measurements.[5,27,28,31-34] In the lightly-doped region, the value of $J_\parallel$ thus obtained is nearly proportional to $T_\mathrm{N}$, and gradually decreases with the increase of $p$, as well. However, in AF-SC samples with $p \cong 0.03$, $J_\parallel$ is still as large as $\sim 0.1$ eV and, therefore, the AF couplings between Cu spins, which will be responsible for the pairing interactions of conduction holes, remains very strong.

The DM coupling constant $J_\mathrm{DM}$, given by eq. (3), is nearly proportional to $J_\parallel$ because $\phi$ and $g$ are little dependent on $p$ and, therefore, also decreases gradually with the increase of $p$. The values of $J_\mathrm{DM}$ evaluated for $p \cong 0$, $0.015$ and $0.03$ from eq. (3) are $\sim 0.97$ meV, $\sim 0.88$ meV and $\sim 0.77$ meV, respectively. Furthermore, the 2D AF spin correlation length at $T_\mathrm{N}$, $\xi_{2\mathrm{D}}(T_\mathrm{N})$, is evaluated from eq. (8) to be $\sim 140$ Å for $p \cong 0$, $0.015$ and $0.03$; it is almost the same in the lightly-doped region.

Table I: Quantities regarding the magnetism of LCO for $p \cong 0$ (LCO-A2), $0.015$ (LCO-B1) and $0.03$ (LCO-C1), and $T_c$ for $p \cong 0.03$ (LCO-C1).

| $p$ | $T_c$/K | $T_N$/K | $M_{WF}(0)/10^{-3}\mu_B$ | $H_c(0)/10^4$Oe | $M_s(0)/\mu_B$ | $\theta/10^{-3}$rad | $J_\parallel$/eV | $J_\perp$/μeV | $J_{DM}$/meV | $\xi_{2D}(T_N)$/Å |
|---|---|---|---|---|---|---|---|---|---|---|
| 0 (LCO-A2) | non-SC | 323 | 3.4 | 6.2 | 0.81 | 4.2 | 0.12 | 4.5 | 0.97 | 140 |
| 0.015 (LCO-B1) | non-SC | 297 | 3.1 | 6.1 | 0.75 | 4.1 | 0.11 | 4.7 | 0.88 | 140 |
| 0.03 (LCO-C1) | 32.5 | 266 | 2.9 | 5.9 | 0.71 | 4.1 | 0.10 | 4.7 | 0.77 | 140 |

The quantities regarding the magnetism of LCO are summarized in Table I for $p \cong 0$, $0.015$ and $0.03$, together with the $T_c$ value for $p \cong 0.03$. Most of these physical quantities are reduced with a small amount of hole-doping, except for $\theta$, $J_\perp$ and $\xi_{2D}(T_N)$. However, it is noteworthy that in AF-SC samples with $p \cong 0.03$, whose AF ordered state is highly homogeneous as in other AF samples, even the largest of their reductions is less than $20\%$ of the value for $p \cong 0$; it is only $\sim 15\%$ for the magnitude of sub-lattice moments originating from $3d_{x^2-y^2}$ electrons localized on Cu sites and the in-plane AF couplings between their spins. Therefore, the strong on-site localization of $\mathrm{Cu}3d_{x^2-y^2}$ electrons still remains in AF-SC samples with $p \cong 0.03$. These findings suggest that the superconductivity whose $T_c$ is as high as $\sim 30$ K is caused by conduction holes with a small density which are hosted in a highly homogeneous and robust AF ordered state of strongly localized $\mathrm{Cu}3d_{x^2-y^2}$ electrons, as has been reported in recent NQR studies.[25)]

### *3.2 Electrical conduction in LCO*

It has been confirmed in the present study that in extremely low doped AF samples with $p \cong 0.015$, the electrical conduction of doped holes has the following three features as reported in previous studies.[27,30,37)] i) The resistivities for both orientations, $\rho_\parallel$ and $\rho_\perp$, increase monotonically with the lowering of $T$, and exhibit no anomalies around $T_N$.[30,37)] ii) The $T$ dependence of $\rho_\parallel$ fits well to a function of $\rho(T) = \rho_0 \exp(T_0/T)^{1/3}$ ($\rho_0$ and $T_0$: constants) at low temperatures below $\sim 100$ K, expected for the 2D VRH of carriers among strongly Anderson-localized states.[30,37,56)] iii) The electrical conduction is very sensitive to the change in magnetic structure due to the field-induced WF transition (see Fig. 5(a)).[27,30)]

Figure 5(a) shows the $H$ dependences of $\rho_\parallel$ and $\rho_\perp$ at low temperatures near $40$ K for a sample (LCO-B2) with $p \cong 0.015$. The $\rho$ data were measured while sweeping $H$ along the out-of-plane direction at a rate of $3$ kOe/min after cooling the sample at zero-field down to each selected temperature. For both the in-plane and out-of-plane directions, $\rho$ drops largely around $H_\mathrm{c}$ with the increase of $H$; its reduction is more than 10 % of the average value around $H = 0$, $\rho(0)$, as reported previously.[27,30] This corroborates that the large enhancement of electrical conductivity with the field-induced WF transition is a characteristic feature for strongly Anderson-localized holes in LCO. A new finding in this study is that the field-induced WF transition in $\rho$-$H$ curves is very broad except for the $\rho_\parallel$-$H$ curve in the process of decreasing $H$; the broadness is different between the processes of increasing and decreasing $H$. Such features for the field-induced WF transition have not been observed in $M$-$H$ curves, which change very sharply around $H_\mathrm{c}$ in both the processes of increasing and decreasing $H$.[30,34,38] It has been suggested that the AF ordered state for $H \lesssim H_\mathrm{c}$ contains a domain structure in which the phase of AF spin alignment is reversed between adjacent domains;[38] such magnetic domains are eliminated by the field-induced WF transition which is accompanied by the flopping of Cu moments with an out-of-plane component antiparallel to the applied magnetic field. Interestingly, a part of holes have been considered to be confined in the antiphase AF-domain boundaries.[38] The present finding for the field-induced WF transition in $\rho$-$H$ curves would suggest an interplay between electric field and/or current and such antiphase AF-domain boundaries in LCO.

In contrast to the robust magnetism, the electrical conduction in lightly-doped AF-SC samples is largely different from that in extremely low doped AF samples. Figure 6 shows the $T$ dependences of $\rho_\parallel$ and $\rho_\perp$ for one piece (LCO-C2) of air-annealed AF-SC samples with the same values of $p$ ($\sim 0.03$) and $T_\mathrm{N}$ (266 K) as for LCO-C1. Although the out-of-plane $\rho_\perp$ increases almost linearly with the lowering of $T$ in a high-$T$ region around 300 K, it starts to deviate downward from a straight line around $T^* \cong 280$ K, where $\chi$ starts to increase steeply, and then decreases largely below a temperature slightly higher than $T_\mathrm{N}$. Such a large change in $\rho_\perp$ is accompanied by a hysteresis with a width of $\sim 4$ K for sample-cooling and heating processes with a rate of 1 K/min. As temperature further decreases, $\rho_\perp$ reaches minimum around 200 K, which is nearly $1/5$ of the maximum around $T^*$, and then increases almost linearly with a very small slope until it meets a small peak just before the SC transition.

On the other hand, the in-plane $\rho_\parallel$ continues to increase even around $T^*$, deviating upward from the higher-temperature very weak $T$ dependence below $\sim T^*$, as shown in Figs. 6 and 7(a). The increase of $\rho_\parallel$ with the lowering of $T$ is comparatively gradual down to $\sim 100$ K, although it becomes a little bit steeper in the $T$ range from $\sim 250$ K to $\sim 200$ K where a hysteresis for sample-cooling and heating processes appears clearly also in $\rho_\parallel(T)$ as seen in Fig.

7(a). In the low-$T$ region below $\sim 100$ K, $\rho_{\parallel}$ increases rapidly by nearly one order of magnitude until it turns to decrease around 50 K due to the development of SC fluctuations.

From the fact that holes doped at extremely low densities are strongly Anderson-localized although Cu-O planes are comparatively clean from the perspective of microscopic crystallinity as mentioned in Sec. 2.1, one can see that their electronic states are very susceptible to microscopic disorders such as impurities and lattice defects. This tempts us to suppose that the low-$T$ in-plane resistivity anomaly for $p \cong 0.03$ would also arise from the strong Anderson localization. As shown in Fig. 8, however, the $T$ dependence of $\rho_{\parallel}$ below $\sim 100$ K for $p \cong 0.03$ cannot be fitted to that in the 2D VRH conduction, differing from the case in the extremely low doping region.[30,37,56] Furthermore, in our studies on several pieces of crystals with $p \cong 0.03$, it has been found that the rapid increase of $\rho_{\parallel}$ below $\sim 100$ K is strongly sample dependent, while the SC $T_c$ does not correlate with the degree of its increase. Since the Anderson localization weakens the screening effect on the Coulomb repulsion between electrons, it reduces their effective attractive interactions and the density of states at the Fermi level generally, leading to a reduction of $T_c$.[57] The superconductivity in LCO will also be suppressed directly by microscopic disorders that give rise to the Anderson localization, because it is considered to be of $d$-wave type.[58] Therefore, given the lack of correlation between the low-$T$ in-plane resistivity anomaly and $T_c$, it is unlikely that this anomaly arises from the Anderson localization of doped holes, which is consistent with the fact that the $T$ dependence of $\rho_{\parallel}$ at low temperatures cannot be explained in terms of the 2D VRH conduction. This finding suggests that for $p \cong 0.03$ AF-SC LCO, doped holes are no longer Anderson-localized. The low-$T$ in-plane resistivity anomaly for $p \cong 0.03$ might be attributed to sample-dependent, macroscopic crystalline imperfections, such as the twin structure in the orthorhombic phase below $T_d$; the twin structure also acts as a magnetic domain structure below $T_N$, since the in-plane components of the sub-lattice moments differ by $90°$ between adjacent domains. However, its origin has not yet been elucidated.

In Fig. 7(b), the anisotropy ratio of resistivity $\rho_{\perp}/\rho_{\parallel}$ is plotted as a function of $T$ for $p \cong 0.03$. As temperature decreases, $\rho_{\perp}/\rho_{\parallel}$ takes a maximum of $\sim 590$ around $T^* \cong 280$ K and then decreases rapidly down to $T \cong 200$ K, where it becomes small by about one order of magnitude compared to the maximum. In the present crystals with $p \cong 0.03$, $\rho_{\parallel}$ does not decrease but increases more rapidly below $\sim T^*$, as mentioned above. Furthermore, the reduction of $\rho_{\perp}$ below $\sim T^*$ in these samples is substantial. Consequently, the anisotropy ratio $\rho_{\perp}/\rho_{\parallel}$ is largely reduced below $\sim T^*$, as shown in Fig. 7(b). Therefore, in the $T$ region below $\sim T^*$, holes doped into Cu-O planes at a very small density are easier to transfer along the out-of-plane direction, and the 2D anisotropy of their electrical conduction becomes weaker. Such a change in the electrical conduction of doped holes might be associated with the rapid development of 2D AF spin correlations below $\sim T^*$, which will be discussed in the next section.

Anomalous behaviors of $\rho_{\parallel}(T)$ and $\rho_{\perp}(T)$ similar to our results were reported in previous studies on LCO single crystals with $p \cong 0.03$,[59] although the $T_{\rm N}$ (250 K) and $T^*$ ($\sim$265 K) values are $\sim$15 K lower than those in our samples with $p \cong 0.03$. Furthermore, it was reported in previous studies that $\rho_{\parallel}$ also decreases below $\sim T^*$ to almost the same degree in the ratio of the minimum around 200 K with respect to the maximum around $T^*$ as in $\rho_{\perp}$.[20,60,61] Although such a discrepancy for $\rho_{\parallel}(T)$ seems to be due to differences in the method of sample preparation such as crystal growth and excess oxygen introduction or problems for the resistivity measurement, for example, whether electrical current flows uniformly through sample between voltage electrodes, the reason is unclear at present.

Another feature for the electrical conduction of doped holes in AF-SC samples with $p \cong 0.03$, differing from that in extremely low doped AF samples whose holes are strongly Anderson-localized, is that the resistivity reduction associated with the field-induced WF transition for $p \cong 0.03$ is much smaller than that for $p \cong 0.015$, as shown in Fig. 5.[27,30] The magnitude of resistivity reduction around $H_{\rm c}$ for $p \cong 0.03$ is only a few percent of the zero-field value $\rho(0)$ in a $T$ region above $\sim$50 K where SC fluctuations are little developed. Furthermore, the field-induced WF transition in $\rho$-$H$ curves occurs rather sharply around $H_{\rm c}$ for $p \cong 0.03$, compared to extremely low doped AF samples. Therefore, the value of $H_{\rm c}$ for $p \cong 0.03$ can be evaluated from $\rho$-$H$ curves in the same manner as for $M$-$H$ curves, and is also plotted in Fig. 4(b), as mentioned in Sec.3.1.

## 4. Discussion

So far, the anomalous behaviors of $\rho_{\parallel}(T)$ and $\rho_{\perp}(T)$ in LCO have been discussed in terms of the phase separation into oxygen/hole-rich and poor domains, and there has been no consensus on the origin. Koga *et al.* suggested from X-ray diffraction experiments and resistivity measurements for $p \cong 0.03$ that the resistive anomalies arise from the appearance of more conductive oxygen/hole-rich domains with a volume fraction of $\sim$5 % and less conductive oxygen/hole-poor domains with a volume fraction of $\sim$95 % due to the phase separation.[59] However, the $T$ range in which the phase separation is underway is much narrower than the $T$ range in which the resistive anomalies appear. A similar discrepancy can be seen for other doping levels, as well.[59,61] Yu *et al*. pointed out that the phase separation is not the direct origin of the resistive anomaly because the former occurs at a temperature higher than the temperature at which the latter occurs.[61] They also reported that structural phase transitions are induced below $\sim 300$ K by the internal elastic strain associated with the existing phase-separated domains, and moreover suggested on the basis of the correlation polaron model proposed by Goodenough *et al*. that the structural phase transitions are accompanied by the anomalous reductions of $\rho_{\parallel}(T)$ and $\rho_{\perp}(T)$ as they observed.[61-63] In the correlation polaron model,[62,63] intermediate-size polarons with a reduced correlation energy, trapping holes, are formed in oxygen/hole-rich domains with shorter covalent $\mathrm{Cu\text{-}O}$ bonds embedded in a oxygen/hole-poor background with longer ionic $\mathrm{Cu\text{-}O}$ bonds, and can move diffusively as the internal strain energy is reduced. Therefore, it was interpreted that the relieving of internal elastic strain due to the structural phase transitions causes the diffusive conduction of polarons, leading to the reduction of resistivity for both the in-plane and out-of-plane directions.[61] However, this is not the case in the present crystals with $p \cong 0.03$, in which $\rho_{\parallel}$ does not decrease but increases more rapidly below $\sim T^{*}$ with the lowering of $T$. Furthermore, the spatial structure of phase separation in the correlation polaron model seems to be inconsistent with that demonstrated in neutron scattering experiments,[23,24] consisting of oxygen/hole-poor domains with an orthorhombic Cmca structure and oxygen/hole-rich domains with a superlattice structure called "stage structure," as mentioned in Sec. 1. In this superlattice structure, the phase for the tilting of $\mathrm{CuO_6}$ octahedra is reversed with a commensurate or incommensurate length along the out-of-plane direction, and excess oxygen atoms are introduced mainly around the boundaries between layers with opposite phases for the octahedral tilting.[23,24]

The lower and upper boundaries of the miscibility gap for the amount of excess oxygen $\delta$ in LCO have not yet been determined definitely,[36,40,59,62] which might be different depending on the method of sample preparation. As already mentioned, it has been demonstrated in NQR studies on the present crystal with $\delta \cong 0.015$ that superconductivity occurs homogeneously, which is not accompanied by the phase separation.[25] This finding indicates that our $\delta \cong 0.015$ crystals lie on the outer low side of the first miscibility gap. In this paper, therefore, we are discussing the resistive anomaly below $\sim T^{*}$ and the associated rapid development of 2D AF spin correlations from a point of view different from the phase separation, that is, as bulk properties for the $\mathrm{Cu\text{-}O}$ plane that exhibits both AF and SC orders simultaneously.

In single-layer cuprate LSCO, the AF ordered phase vanishes at a very small hole-doping level of $p \cong 0.02$,[7,44] while in the same type of single-layer cuprate LCO, $T_{\mathrm{N}}$ is higher than $250\ \mathrm{K}$ even for $p \cong 0.03$. Furthermore, in multi-layered cuprates, in which doped holes are easier to transfer between Cu-O planes within the unit layer, the AF ordered phase extends up to a larger hole-doping level over $p \cong 0.06$.[18] The strong suppression of AF ordering due to doping holes with the substitution of Sr for La in LSCO is considered to arise from the following two effects: disorder effects due to the Sr substitution on the microscopic crystallinity and disturbance effects due to the in-plane motion of doped holes on the 2D AF spin correlations. According to the mean-field theory on the Cu spin system of LCO,[34] $T_{\mathrm{N}}$ is proportional to $(\xi_{\mathrm{2D}}(T_{\mathrm{N}})/a_0)^2$ or the number of Cu spins included within an area of $(\xi_{\mathrm{2D}}(T_{\mathrm{N}}))^2$ in Cu-O planes, as shown in eq. (7), and the factor $4\theta J_{\mathrm{DM}} + J_{\perp}$ corresponds to the energy gain per Cu spin due to the 3D AF ordering. Microscopic crystallographic disorders due to the Sr substitution in LSCO, as observed in La-NQR measurements,[43] will affect the physical quantities such as magnetic couplings in eq. (7), consequently reducing $T_{\mathrm{N}}$. As holes are doped into Cu-O planes, they mainly occupy the in-plane $\mathrm{O}2p_{x(y)}$ and $\mathrm{Cu}3d_{x^2-y^2}$ orbitals of $\mathrm{CuO_6}$ octahedra,[8,64] which are involved in the superexchange interactions, and break the AF couplings between their surrounding Cu spins. Therefore, the development of 2D AF spin correlations is disturbed by the in-plane motion of doped holes through these orbitals,[8,64,65] which contributes to the reduction of $T_{\mathrm{N}}$ significantly. In fact, it has been revealed in neutron inelastic scattering experiments on LSCO that the growth of $\xi_{\mathrm{2D}}$ at low temperatures is largely suppressed with the increase of $p$ and limited to order $10\ \mathrm{Å}$ at high doping levels around $p \cong p_{\mathrm{o}}\ (\sim 0.16)$.[66,67]

On the other hand, in $p \cong 0.03$ LCO with relatively clean Cu-O planes, the growth of $\xi_{\mathrm{2D}}$ accelerates below $T^* \cong 280\ \mathrm{K}$ and reaches to a large extent enough to cause the 3D AF ordering at $T_{\mathrm{N}} = 266\ \mathrm{K}$. In a $T$-region below $\sim T^*$, $\rho_{\perp}$ decreases largely while $\rho_{\parallel}$ increases more rapidly with the lowering of $T$; the 2D anisotropy of electrical conduction in this system becomes weak significantly, as mentioned in Sec. 3.2. To understand the origin of such anomalous behaviors in the magnetism and electrical conduction, we revisit the electronic orbitals occupied by doped holes. Theoretical studies for the electronic structure in La-based high-$T_{\mathrm{c}}$ cuprates have pointed out that the apical $\mathrm{O}2p_z$ and $\mathrm{Cu}3d_{3z^2-r^2}$ orbitals extending along the out-of-plane direction in $\mathrm{CuO_6}$ octahedra also contribute to the low-energy states around the Fermi level.[68,69] The degree of their contributions to the low-energy states with respect to the leading orbitals, in-plane $\mathrm{O}2p_{x(y)}$ and $\mathrm{Cu}3d_{x^2-y^2}$, depends on the ratio of apical O-Cu distance to in-plane O-Cu distance or the Jahn-Teller distortion ratio.[68,69] Interestingly, it has been reported in X-ray diffraction, thermal expansion coefficient and heat capacity measurements on LCO that a lattice distortion also occurs around $T^*$.[59,61] If this lattice distortion enhances the contributions of apical $\mathrm{O}2p_z$ and $\mathrm{Cu}3d_{3z^2-r^2}$ orbitals to the low-energy states or increases the occupation rates of doped holes in these orbitals and, conversely, decreases those in in-plane $\mathrm{O}2p_{x(y)}$ and $\mathrm{Cu}3d_{x^2-y^2}$ orbitals, this could qualitatively explain that the 2D anisotropy in the electrical conduction of doped holes becomes weak below $\sim T^*$. Furthermore, the reduction in the occupation rates of doped holes in in-plane $\mathrm{O}2p_{x(y)}$ and $\mathrm{Cu}3d_{x^2-y^2}$ orbitals below $\sim T^*$ could also accelerate the growth of $\xi_{\mathrm{2D}}$ around this temperature, because it is expected to weaken the disturbing effect on 2D AF spin correlations due to the in-plane motion of doped holes through these orbitals. Thus, the anomalous behaviors below $\sim T^*$ in the magnetism and electrical conduction of $p \cong 0.03$ LCO could be explained qualitatively in terms of the idea that the lattice distortion is accompanied by a change in the occupation rates of doped holes in electronic orbitals. However, whether such a lattice distortion actually occurs around $\sim T^*$ in $p \cong 0.03$ LCO is unclear at present.

From evaluations on the quantities such as $M_{\mathrm{s}}$ and $J_{\parallel}$ regarding the magnetism of LCO for the three doping levels, it has been suggested that the localization of $3d_{x^2-y^2}$ electrons on Cu sites remains very strong even for $p \cong 0.03$. In the AF-SC state for $p \cong 0.03$, the sub-lattice moments on Cu sites and their AF couplings still remain large, almost unchanged from those of the parent material that is a strongly correlated Mott-insulator, and the average distance $\bar{l}$ between conduction holes in Cu-O planes, evaluated from $\bar{l} = a_0/\sqrt{p}$, is as large as $\sim 22$ Å, which is comparable to the SC coherence length or the characteristic length for pairing interactions. Thus, the electronic system in AF-SC LCO with $p \cong 0.03$ is quite unique for a superconductor, which tempts us to suppose that the mechanism for this superconductivity, including the pairing interactions of doped holes and the SC phase transition mechanism, might be novel and highly interesting.

## 5. Summary

We have studied the magnetism and electrical conduction of single-layer cuprate LCO comparing those between AF-SC samples with $p \cong 0.03$ and AF samples with $p \cong 0$ and $0.015$, and obtained interesting results revealing their intrinsic properties in the lightly-doped Mott insulator. In AF-SC samples with $p \cong 0.03$, it has been found from the $T$ dependences of $\chi$ and $M$-$H$ curve that 2D AF spin correlations develops very gradually at high temperatures, but their development accelerates rapidly below $T^* \cong 280$ K, leading to the subsequent 3D AF ordering at $T_{\mathrm{N}} = 266$ K. The superconductivity occurs at $T_{\mathrm{c}} = 32.5$ K in the homogeneous and robust AF ordered state that retains sizable sub-lattice moments ($M_{\mathrm{s}} \cong 0.7\mu_{\mathrm{B}}$) and in-plane AF couplings ($J_{\parallel} \cong 0.1$ eV). Measurements on the $T$ dependence of resistivity have suggested that conduction holes in lightly-doped AF-SC samples with $p \cong 0.03$ are no longer strongly Anderson-localized, unlike those in extremely low doped AF samples with $p \cong 0.015$. Furthermore, it has been demonstrated for $p \cong 0.03$ that in a $T$-region below $\sim T^{*}$, $\rho_{\perp}$ decreases largely, while $\rho_{\parallel}$ increases more rapidly; the anisotropy ratio $\rho_{\perp}/\rho_{\parallel}$ decreases significantly. The anomalous behaviors below $\sim T^{*}$ in the magnetism and electrical conduction of lightly-doped AF-SC LCO seem to arise from the same origin. Further studies are needed to unveil the underlying mechanism.

**Acknowledgment**

The authors would like to acknowledge Y. Matsushita for performing X-ray diffraction measurements on LCO crystals with $p \cong 0.03$ and providing results on the lattice constants, K. Ishida and Y. Uno for supporting NQR measurements and valuable discussions, and M. Iwamatsu and K. Tachi for supporting resistivity measurements and sample preparations. This work is partially supported by the JSPS Grant-in-Aid for Scientific Research (Grant Nos. 24K06950 and 24H01599).

## References

*E-mail: moda@sci.hokudai.ac.jp

**Figures and Figure Captions**

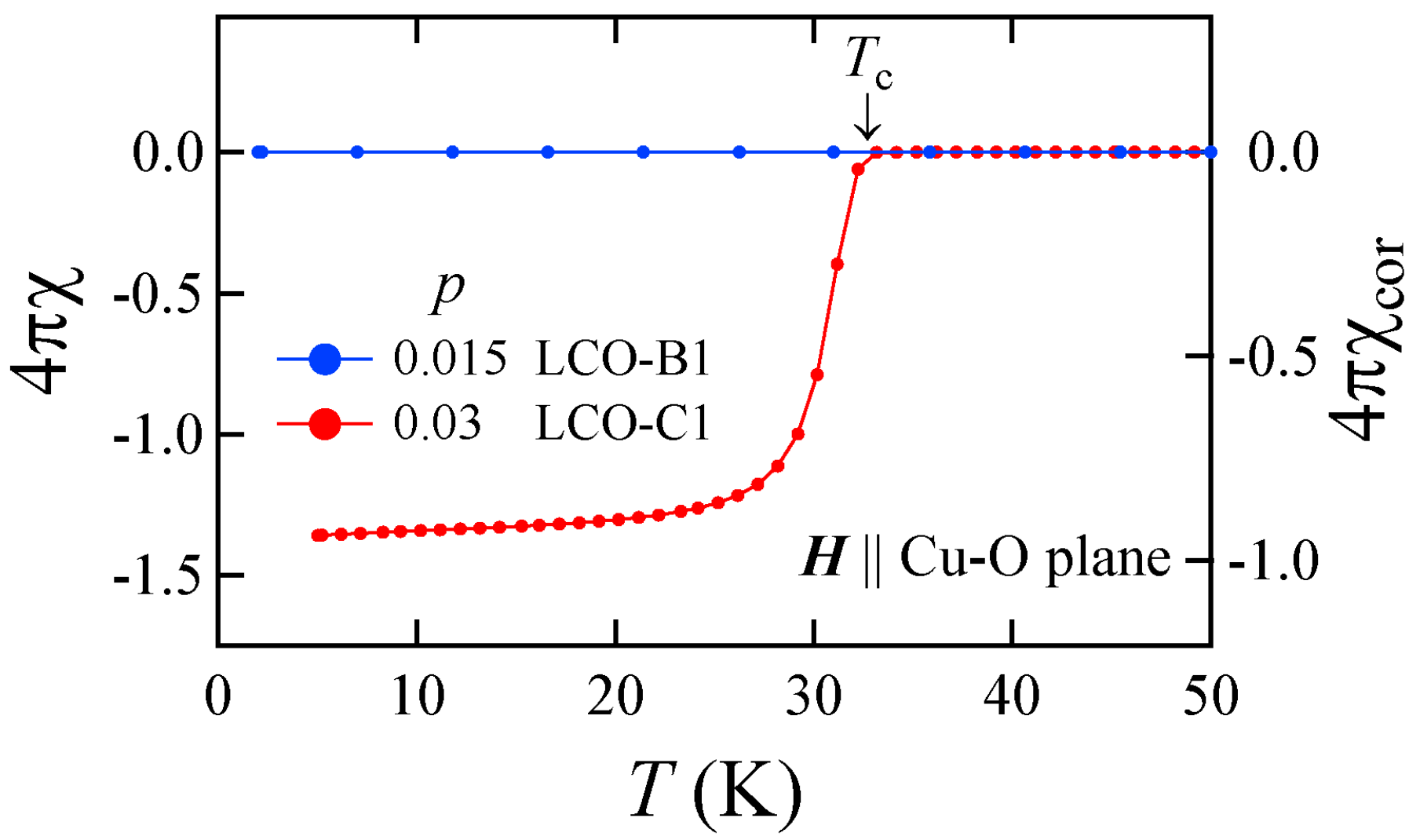

Fig. 1: $T$ dependences of magnetic susceptibility $\chi\ (\equiv M/H)$ multiplied by $4\pi$ ($4\pi\chi$) below 50 K for Ar-annealed $p \cong 0.015$ (LCO-B1) and air-annealed $p \cong 0.03$ (LCO-C1) samples. The measurements of magnetization $M$ were done in the process of rising temperature under a magnetic field $H$ of $2.0$ Oe parallel to Cu-O planes after cooling the sample at zero-field slowly from 300 K to the lowest temperature examined and then applying the magnetic field. The right axis is for the diamagnetic susceptibility $4\pi\chi_{\mathrm{cor}}$ of LCO-C1 that is corrected for the demagnetizing field effect.

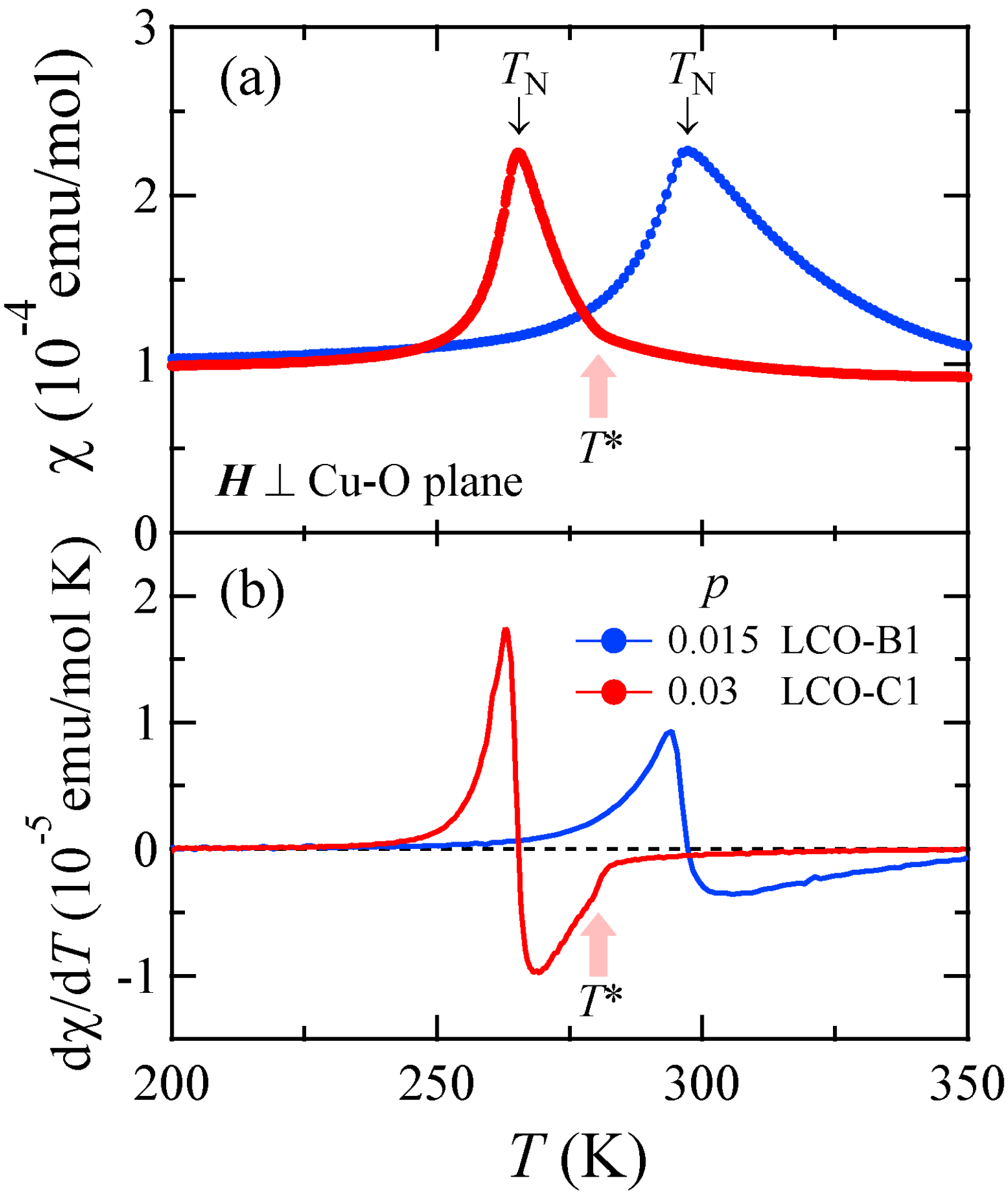

Fig. 2: Magnetic susceptibility $\chi\ (\equiv M/H)$ peak around $T_{\mathrm{N}}$ in $p \cong 0.015$ (LCO-B1) and $p \cong 0.03$ (LCO-C1) samples. (a) $\chi$ vs. $T$ plots and (b) $d\chi/dT$ vs. $T$ plots. The data used here were taken under a magnetic field of $H = 10$ kOe applied along the direction perpendicular to Cu-O planes.

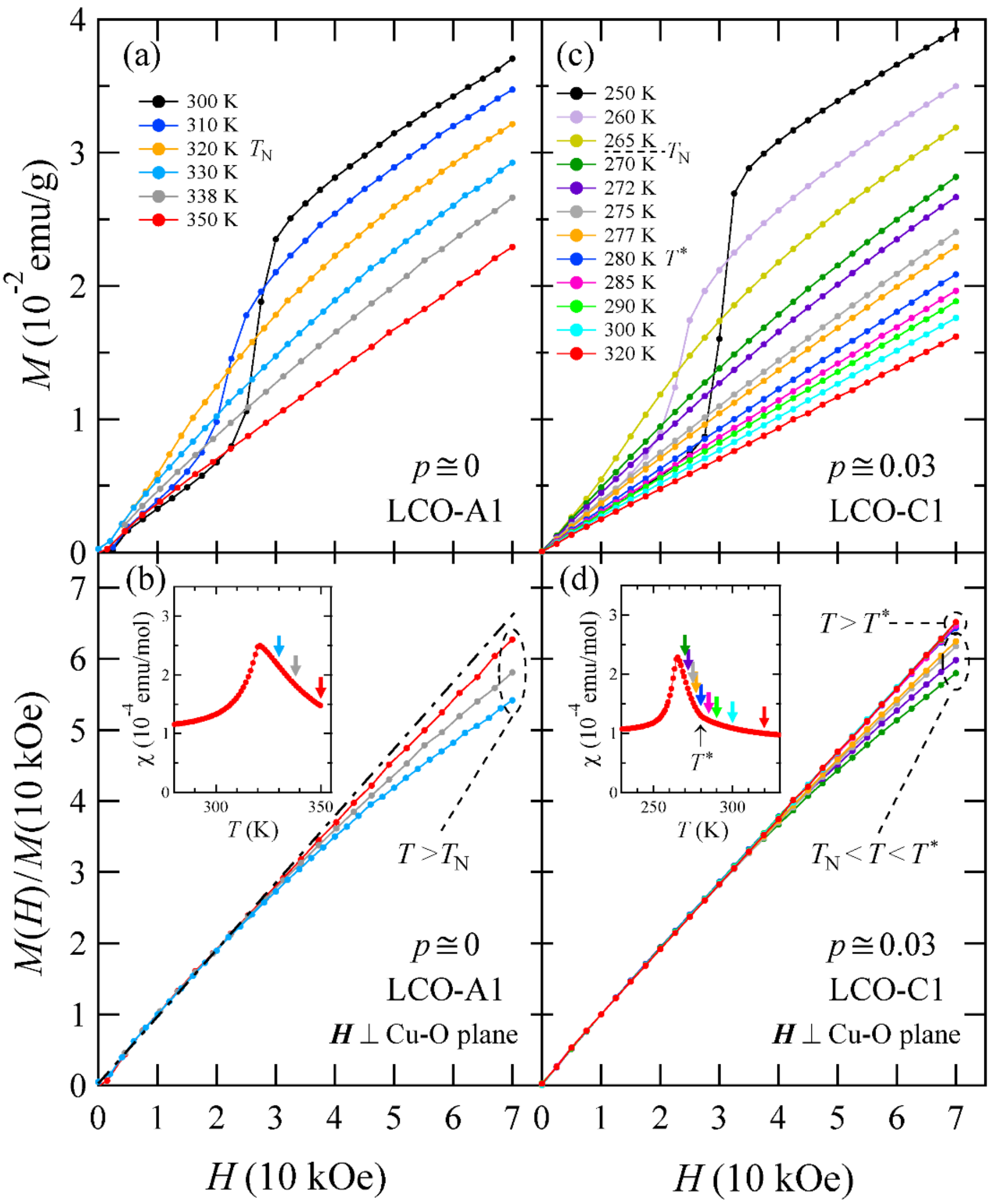

Fig. 3: Non-linear $M$-$H$ curves around $T_{\mathrm{N}}$ for $p \cong 0$ (LCO-A1) and $p \cong 0.03$ (LCO-C1) samples, examined under magnetic fields perpendicular to Cu-O planes. (a) $M$ vs. $H$ plots at $300\ \mathrm{K} \leq T \leq 350\ \mathrm{K}$ for $p \cong 0$ and (c) those at $250\ \mathrm{K} \leq T \leq 320\ \mathrm{K}$ for $p \cong 0.03$. (b) $M(H)/M(10\ \mathrm{kOe})$ vs. $H$ plots at $T > T_{\mathrm{N}}$ for $p \cong 0$ and (d) those at $T > T_{\mathrm{N}}$ for $p \cong 0.03$. In these $M$-$H$ curves, no hysteresis is observed between the field-increasing and decreasing processes. The insets in (b) and (d) show the $T$ dependences of $\chi$ around $T_{\mathrm{N}}$ for $p \cong 0$ and $p \cong 0.03$, respectively. In these insets, the thick colored arrows indicate the temperatures at which the corresponding $M$-$H$ curves were measured. The dash-dotted line in (c) represents a straight line fitted to the data at low fields.

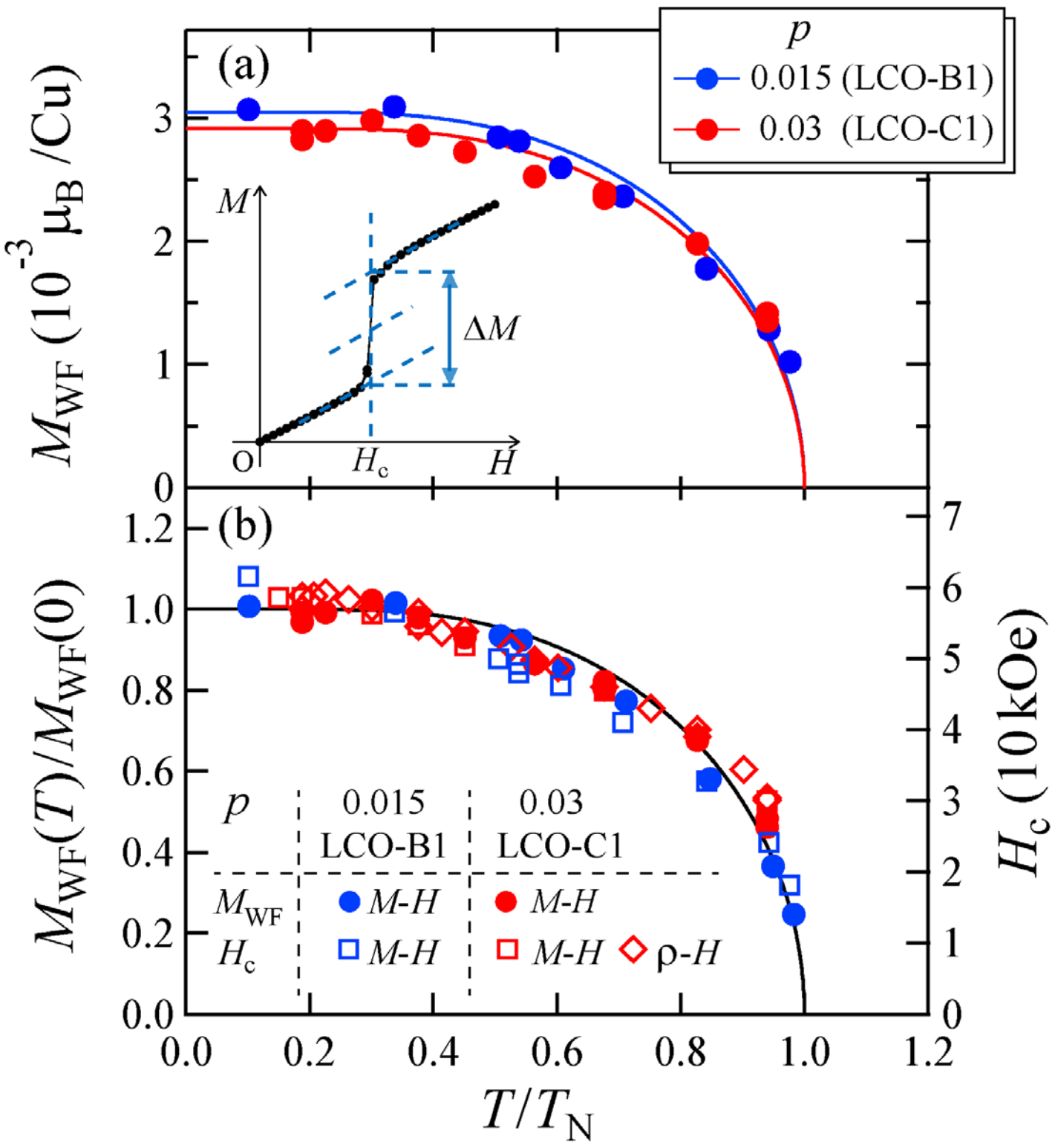

Fig. 4: Weak ferromagnetism in the AF ordered states of $p \cong 0.015$ (LCO-B1) and $p \cong 0.03$ (LCO-C1) samples. (a) $M_{WF}$ vs. $T/T_N$ plots and (b) $M_{WF}(T)/M_{WF}(0)$ and $H_c$ vs. $T/T_N$ plots. The value of WF moment per Cu site $M_{WF}$ was obtained from the jump of $M$ at the critical field $H_c$ of field-induced WF transition by using the relation $M_{WF} = \Delta M/N$, where $\Delta M$ and $N$ are the magnitude of magnetization jump and the number of sub-lattice moments both of which are expressed per unit mass of sample, respectively. The inset shows the way to evaluate $\Delta M$ and $H_c$ in the $M$-$H$ curve. In both lower and higher sides than $H_c$, the $M$-$H$ curve fits to straight lines in parallel with each other, which enables us to evaluate $\Delta M$ as the interval along the $M$-axis between the two fitted parallel lines. On the other hand, we define $H_c$ as the $H$ coordinate of the point at which the center line between the two fitted parallel lines intersects the $M$-$H$ curve jumping with the field-induced WF transition. At temperatures lower than $\sim$200 K, the $M$-$H$ curve exhibits a hysteresis around $H_c$ for the sweeping of $H$. In such cases, we define $H_c$ as the average of the values in the two processes of increasing and decreasing $H$. Furthermore, the field-induced WF transition is accompanied by a resistivity reduction, as shown in the $H$ dependences of $\rho_{\parallel}$ and $\rho_{\perp}$ in Fig. 5. For $p \cong 0.03$, where the resistivity reduction around $H_c$ is comparatively sharp, $H_c$ is also evaluated in the $\rho$-$H$ curves in the same manner as in the $M$-$H$ curves, and its value is plotted in Fig. 4(b). The solid lines show the $T$ dependences of $M_S$ in the mean-field approximation for $S = 1/2$ Heisenberg antiferromagnets, given by the Brillouin function for $S = 1/2$, where $M_S$ is scaled so as to agree with the low-$T$ data of $M_{WF}$ for each sample in (a) and become unity at $T \ll T_N$ in (b).

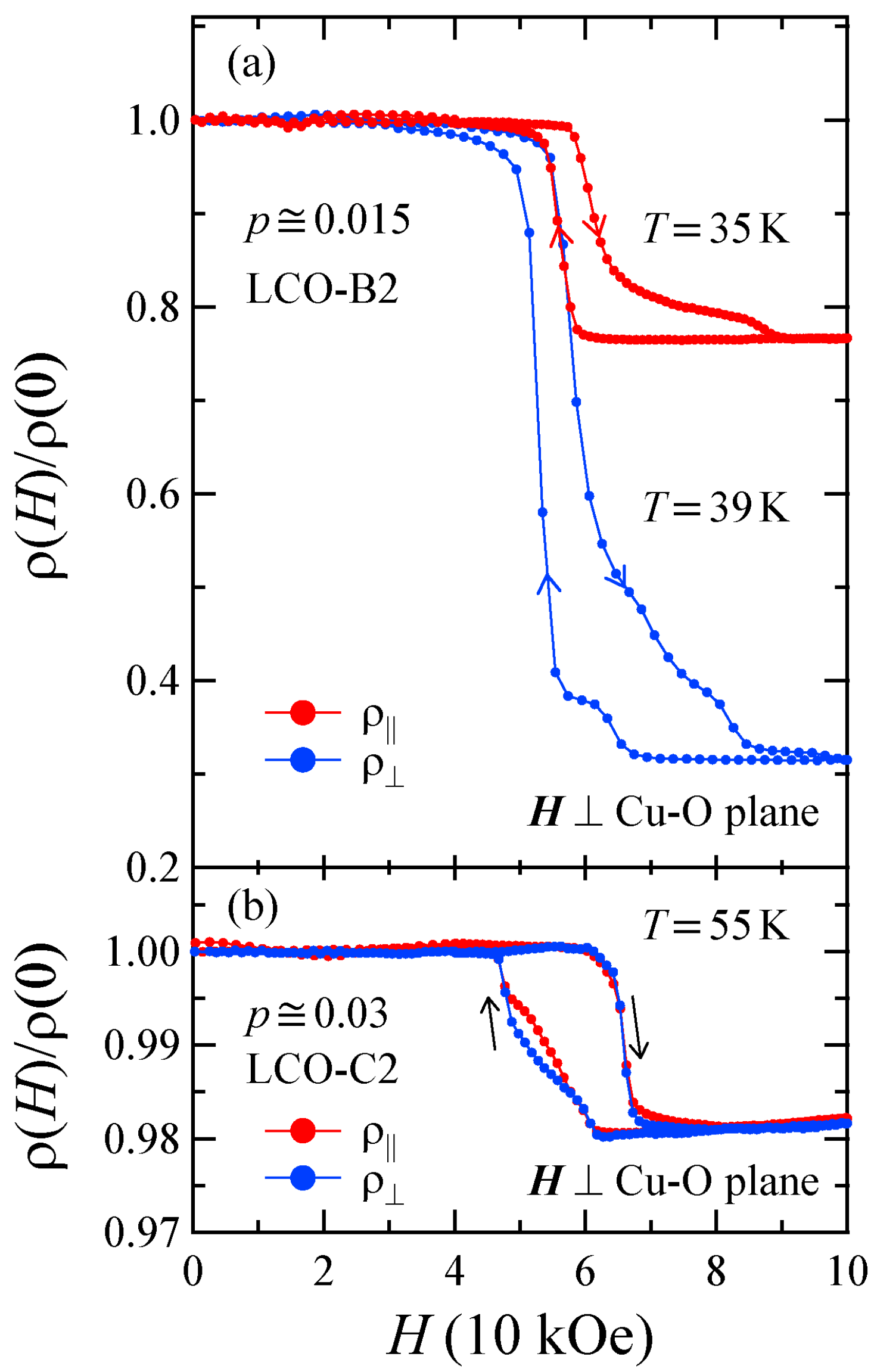

Fig. 5: Resistivity reduction associated with the field-induced WF transition. (a) for $p \cong 0.015$ (LCO-B2) and (b) for $p \cong 0.03$ (LCO-C2). The $\rho$ data were measured while sweeping $H$ along the out-of-plane direction at a rate of $3\ \mathrm{kOe/min}$ after cooling the sample at zero-field down to each selected temperature. The arrows indicate the direction of resistivity change with increasing and decreasing $H$.

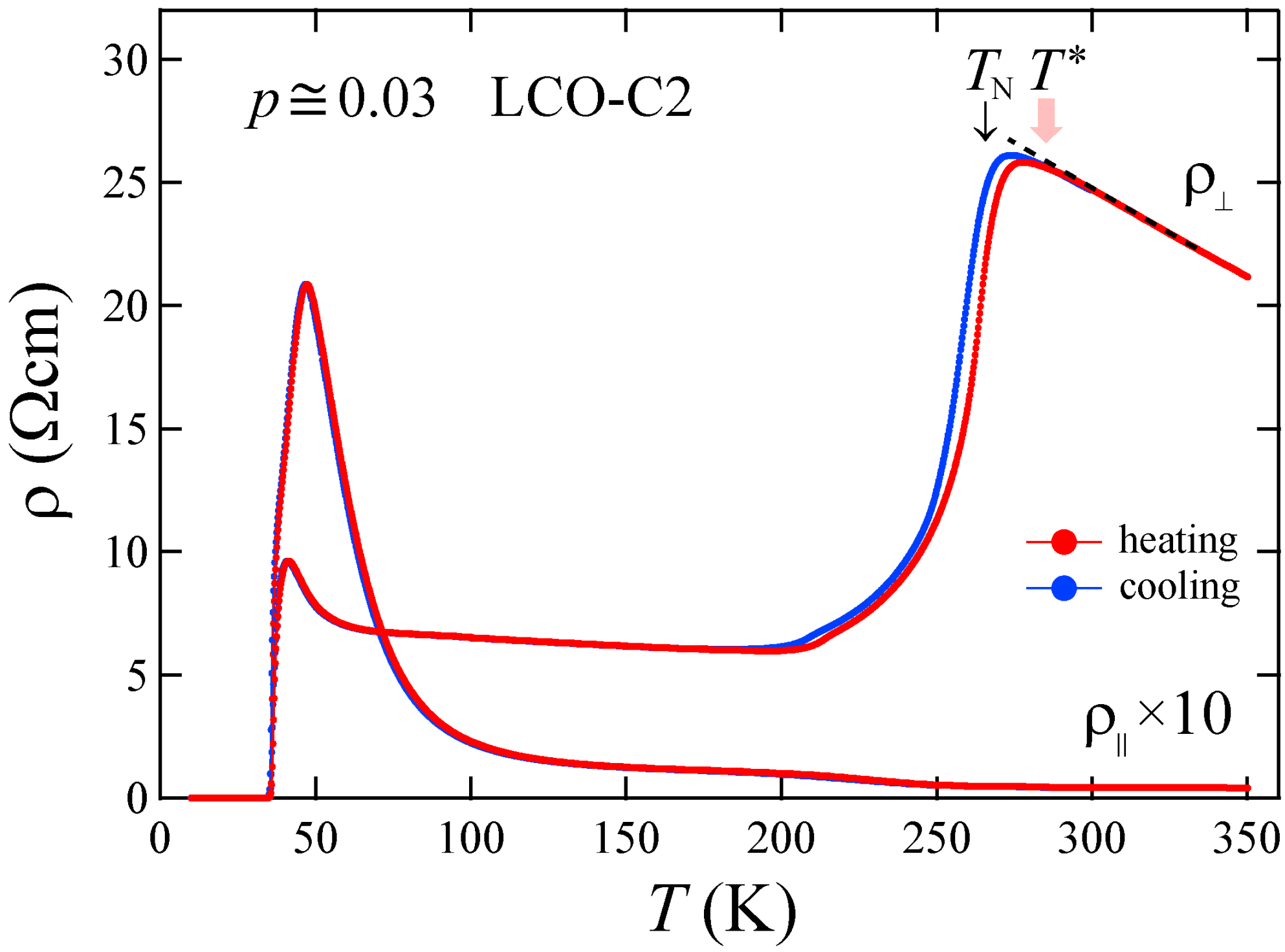

Fig. 6: Anomalous behaviors for the electrical conduction of doped holes for $p \cong 0.03$ (LCO-C2), which can be seen in the $T$ dependences of in-plane and out-of-plane resistivities $\rho_\parallel$ and $\rho_\perp$. The $\rho$ data were obtained for both the sample-heating and cooling processes at a rate of 1 K/min. The broken line shows a linear extrapolation of the high-$T$ $\rho_\perp$ data for $T \gtrsim T^*$ towards lower temperatures.

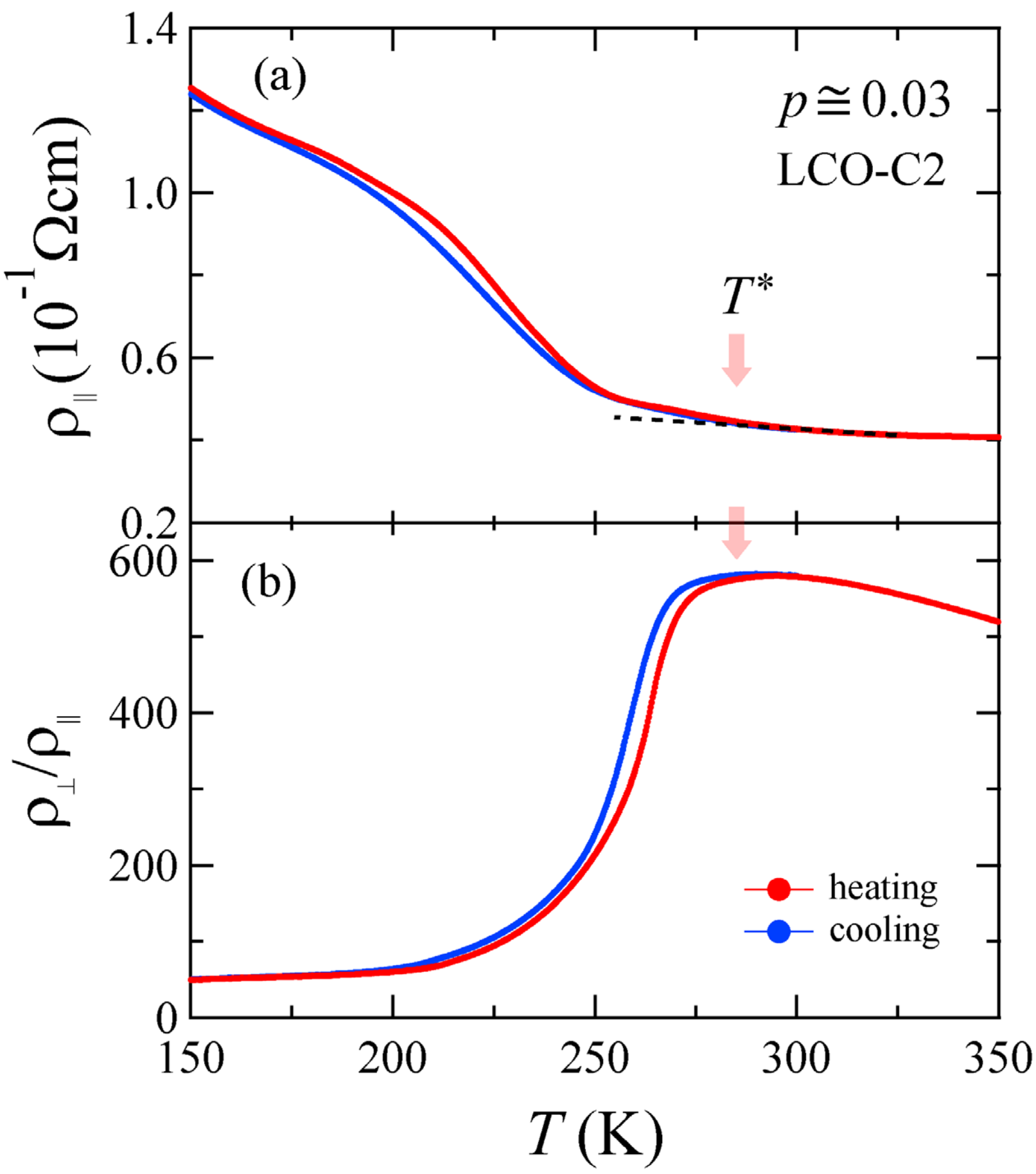

Fig. 7: Anomalous behaviors for the electrical conduction of doped holes for $p \cong 0.03$ (LCO-C2), which can be seen in the $T$ dependences of (a) enlarged $\rho_\parallel$ and (b) anisotropy ratio $\rho_\perp/\rho_\parallel$ for $150\text{K} \leq T \leq 350\text{K}$. The broken line in (a) shows a linear extrapolation of the high-$T$ $\rho_\parallel$ data for $T \gtrsim T^*$ towards lower temperatures.

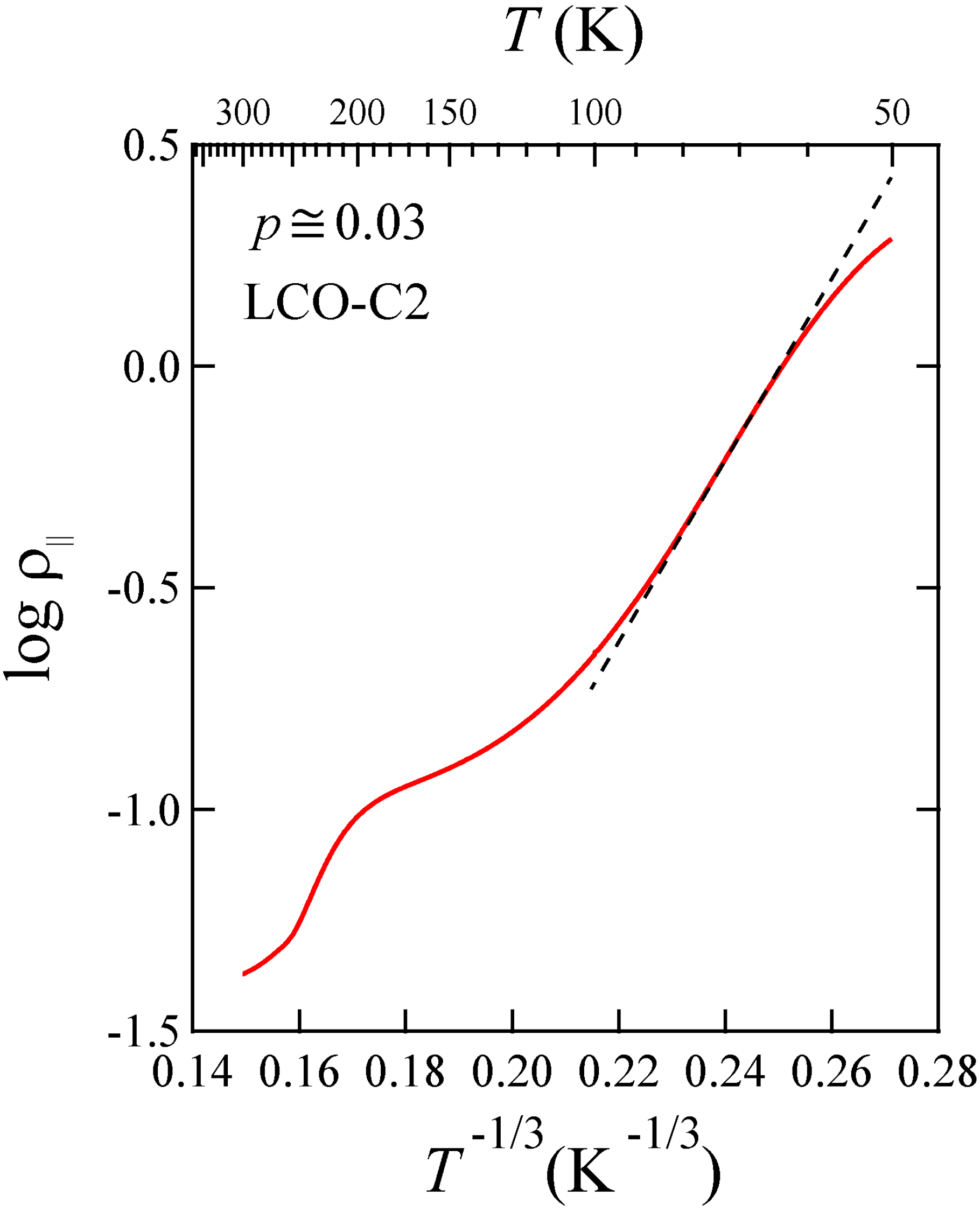

Fig. 8: $\log\rho_{\parallel}$ vs. $T^{-\frac{1}{3}}$ plots for $p \cong 0.03$ (LCO-C2). The data obtained during the sample-cooling process are used here. The straight broken line shows a function of $\rho(T) = \rho_0 \exp(T_0/T)^{1/3}$, which is drawn to make a comparison between the in-plane conduction of doped holes below $\sim 100$ K and the 2D VRH conduction. The $T$ range in which the $\log\rho_{\parallel}$ vs. $T^{-\frac{1}{3}}$ curve can be considered linear is very narrow (from $\sim 65$ K to $\sim 80$ K at most).